# Pseudogap-driven Hall effect sign reversal

D. V. Evtushinsky,[1] S. V. Borisenko,[1] A. A. Kordyuk,[1,2] V. B. Zabolotnyy,[1]
D. S. Inosov,[1] B. Büchner,[1] H. Berger,[3] L. Patthey,[4] and R. Follath[5]

[1]*Institute for Solid State Research, IFW Dresden, P.O. Box 270116, D-01171 Dresden, Germany.*
[2]*Institute of Metal Physics of National Academy of Sciences of Ukraine, 03142 Kyiv, Ukraine.*
[3]*Institut de Physique Appliquée, Ecole Politechnique Féderale de Lausanne, CH-1015 Lausanne, Switzerland.*
[4]*Swiss Light Source, Paul Scherrer Institut, CH-5234 Villigen, Switzerland*
[5]*BESSY GmbH, Albert-Einstein-Strasse 15, 12489 Berlin, Germany.*

We present a calculation of the Hall coefficient in 2H-TaSe$_2$ and 2H-Cu$_{0.2}$NbS$_2$ relied on the photoemission data and compare the results to transport observations. The approach, based on the solution of the semiclassical Boltzmann equation in the isotropic $\tau$-approximation yields high-temperature Hall coefficient consistent with the one measured directly. Taking into account the opening of the pseudogap and redistribution of the spectral weight, recently observed in angle resolved photoemission spectra of 2H-TaSe$_2$, allows us to reproduce the temperature dependence of the Hall coefficient including prominent sign change with *no* adjustable parameters.

Transition metal dichalcogenides, probably the most studied charge density wave (CDW) bearing compounds, exhibit Hall coefficient sign change from hole-like to electron-like soon after the transition into CDW state [1, 2]. Recently the same effect was discovered in high temperature superconductors (HTSC) [3]. Up to the moment reasons for such a temperature behaviour of the transversal conductivity in both types of compounds were not clear, though it was obvious that strong changes in the electronic structure are involved. Another common feature of CDW and HTSC compounds is presence of the pseudogap in excitation spectra [4–7]. Below we show that (i) it is possible to calculate Hall coefficient of a solid based purely on the photoemission data, and (ii) the opening of the anisotropic pseudogap is the main reason for reversion of the Hall effect in the renown CDW system, 2H-TaSe$_2$.

Hall coefficient is very sensitive to the changes of the electronic structure, and therefore undergoes significant modification upon the reconstruction of the Fermi surface, which is often preceded by the formation of the pseudogap. Since charge dynamics in the crystal is restricted to a narrow energy range around the Fermi level, opening of the pseudogap reduces effective number of charge carriers, which should definitely affect transport properties of the solid. In this communication we investigate quantitatively an effect that the opening of the pseudogap causes to conductive properties and suggest a procedure of Hall coefficient calculation from angle resolved photoemission spectroscopy (ARPES) data [8]. Our procedure accounts for the opening of the pseudogap and nonuniform distribution of spectral weight. The actual temperature-dependent calculation of the Hall coefficient is performed for two transition metal dichalcogenides — 2H-TaSe$_2$ and 2H-Cu$_{0.2}$NbS$_2$. TaSe$_2$ undergoes a transition to incommensurate CDW state at 122 K and to commensurate one at 90 K. In contrast, NbS$_2$ exhibits no CDW, and its Hall coefficient slightly depends on temperature, which suggested the choice for the second compound in the calculations.

In our calculations we assume electric field **E** to be parallel to $ab$ plane, and magnetic field **B** parallel to $c$ axis (thus, current **j** flows in $ab$ plane). Such an experimental geometry is common for investigation of two-dimensional compounds, in particular dichalcogenides [1, 2, 9, 10]. In the low-field limit current density is related to electrical field by means of conductivity tensor:

$$\mathbf{j} = \sigma \mathbf{E}, \quad \sigma = \begin{pmatrix} \sigma_{xx} & \sigma_{xy} \\ -\sigma_{xy} & \sigma_{xx} \end{pmatrix}. \quad (1)$$

Components of the conductivity tensor are derived from the solution of semiclassical Boltzmann equation. Neglecting $k_z$ dispersion and taking into account identity of $a$ and $b$ axes, $\sigma_{xx}$ and $\sigma_{xy}$ are expressed through the integrals over the Fermi surface in the first Brillouin zone (formulae given in SI units):

$$\sigma_{xx} = \frac{e^2}{2\pi L_c h} \int \tau(\mathbf{k}) v_F(\mathbf{k}) \, dk \quad (2)$$

$$\sigma_{xy} = \frac{e^3 B}{L_c h^2} \int \frac{\tau^2(\mathbf{k}) v_F^2(\mathbf{k})}{\rho(\mathbf{k})} \, dk, \quad (3)$$

where $\tau$ is a quasiparticle lifetime, $v_F$ — the *renormalized* Fermi velocity, $\rho$ — Fermi surface curvature radius, $dk$ — element of Fermi surface length, $L_c$ — size of elementary cell along the $c$ axis, $h$ — the Plank's constant, $e$ — the elementary charge. Mathematically equivalent, but less convenient for our discussion formulae, may be found in literature [11–13]. By definition the Hall coefficient equals to the Hall electrical field over the magnetic field and the current density: $R_H \equiv E_H/(B \cdot j)$. In terms of conductivity tensor it is expressed in the following way:

$$R_H = \frac{\sigma_{xy}}{B \cdot \sigma_{xx}^2} \quad (4)$$

ARPES gives us complete knowledge about the band structure, therefore the only thing we are missing to calculate the conductivity tensor from Eqs. (2) and (3) is transport

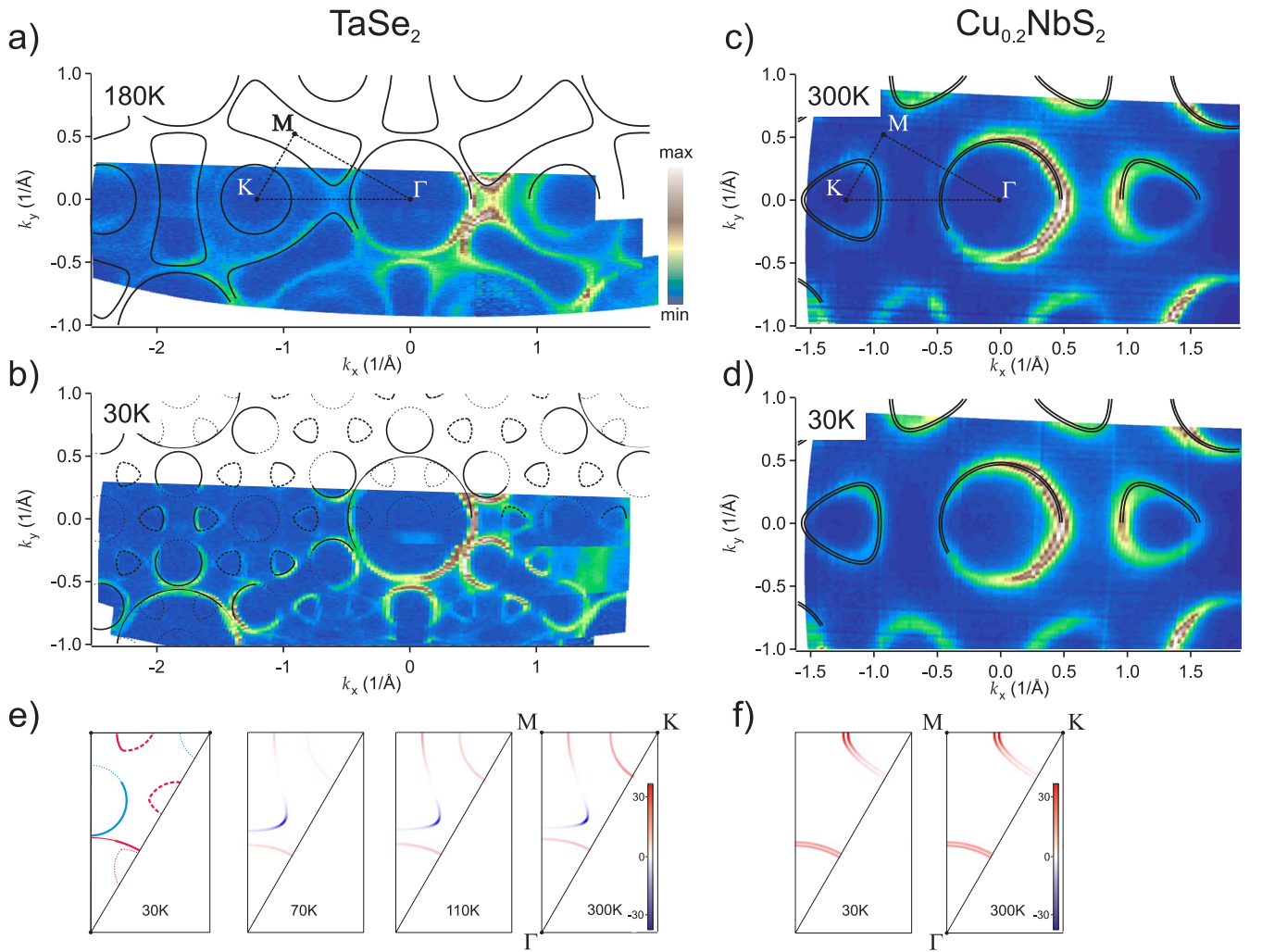

Fig. 1 (color online). Evolution of the Fermi surface of TaSe$_2$ and Cu$_{0.2}$NbS$_2$ with temperature. Fermi surface of TaSe$_2$ changes topology with cooling (a, b). The first signature of Fermi surface reconstruction is the opening of the pseudogap. In normal state tight-binding fit shown (a,c,d). For CDW–reconstructed Fermi surface (b) different types of guidelines correspond to spectral weight distribution. Absence of changes and uniform spectral weight distribution in the spectra of Cu$_{0.2}$NbS$_2$ (c, d). Relative contribution to $\sigma_{xy}$ from different parts of Fermi surface at different temperatures, shown in irreducible part of the Brillouin zone for TaSe$_2$ (e) and Cu$_{0.2}$NbS$_2$ (f).

lifetime $\tau$, which should not be mixed with the quantum lifetime $\tau_q$, seen in ARPES [14, 15]. If we assume that $\tau$ is momentum independent $\tau(\mathbf{k}) = \mathrm{const}$, then $\tau$ cancels out, and the expression for $R_H$ reduces to:

$$R_H = \frac{4\pi^2 L_c}{e} \cdot \frac{\int v_F^2(\mathbf{k})/\rho(\mathbf{k})\,dk}{\left(\int v_F(\mathbf{k})\,dk\right)^2}, \quad (5)$$

where $R_H = \mathrm{m}^3/\mathrm{C}$ [16]. For 2H-TaSe$_2$ $L_c = 12.7$ Å [17], all other quantities that enter Eq. (5) can be straightforwardly extracted from ARPES spectra as they are seen in the ARPES images [15, 18] and the Fermi surface maps [19] per se.

It is well known that for conventional metals Eq. (5) yields a result consistent with direct measurements of $R_H$ [20]. We find that it also gives reasonably good agreement with experiment for NbS$_2$. For TaSe$_2$ Eq. (5) provides correct $R_H$ only at high temperatures and becomes inapplicable at lower ones, when the Fermi surface reconstruction sets on. Formulae (2), (3), and, hence, (5) imply that all energy bands are *equally and uniformly populated* with electrons. Though this assumption often holds, a *complex picture of spectral weight distribution* does not appear to be a rare occasion for unconventional materials [7, 22]. For such a case Eq. (5) should be modified introducing a factor $D(\mathbf{k})$ that takes into account distribution of the spectral weight, i.e. behaviour of density of states (DOS) near the Fermi level:

$$R_H = \frac{4\pi^2 L_c}{e} \cdot \frac{\int D(\mathbf{k}) v_F^2(\mathbf{k})/\rho(\mathbf{k})\,dk}{\left(\int D(\mathbf{k}) v_F(\mathbf{k})\,dk\right)^2}, \quad (6)$$

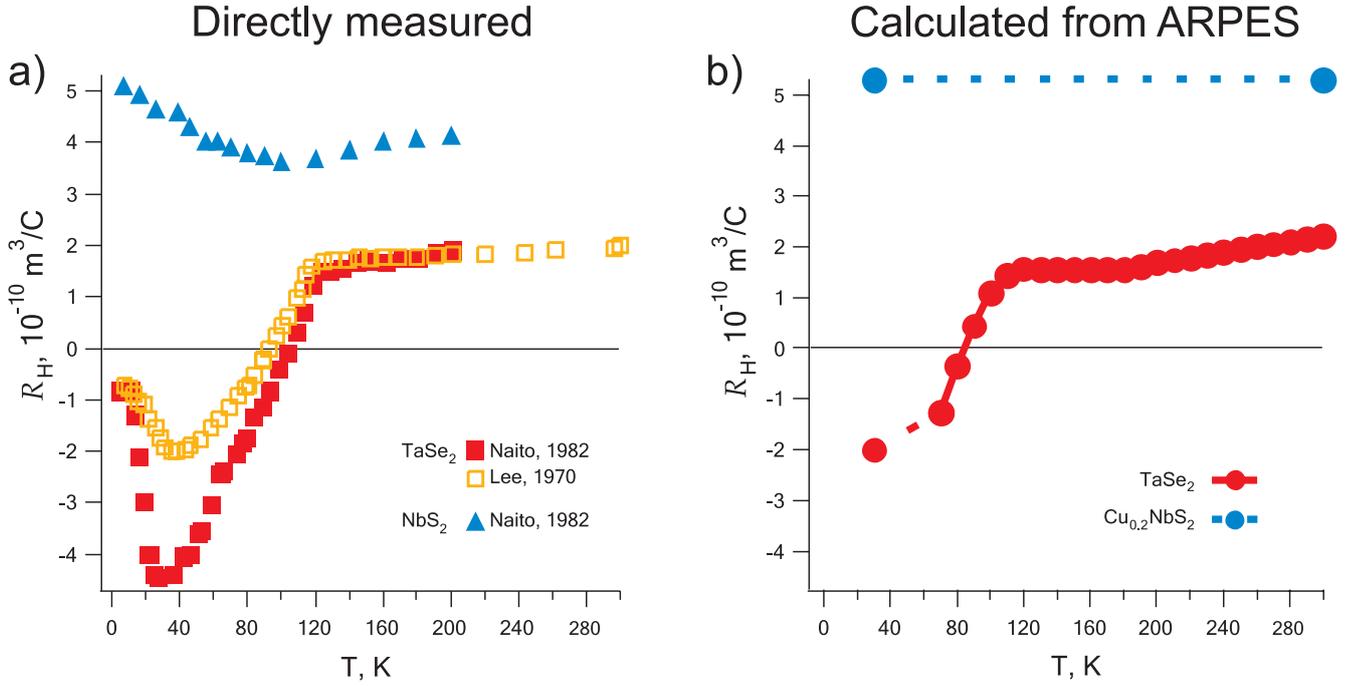

Fig. 2 (color online). Temperature dependence of the Hall coefficient. Hall coefficient in $NbS_2$ has weak temperature dependence, while in $TaSe_2$ Hall effect changes sign (a). Discrepancy between two experimental curves for $TaSe_2$ is due to charge density wave suppression by impurities. Hall coefficient of $Cu_{0.2}NbS_2$, calculated in approximation of equally "populated" bands, agrees well with directly measured, while in case of $TaSe_2$ one should take into account spectral weight redistribution and the opening of the pseudogap (b).

where

$$D(\mathbf{k}) = \int_{-\infty}^{\infty} \text{DOS}_{\mathbf{k}}(\omega) \cdot \left(-\frac{\partial f(\omega)}{\partial \omega}\right) d\omega, \quad (7)$$

that is the temperature weighted DOS at the Fermi level, $f(\omega)$ is the Fermi function, and $\omega$ is the binding energy. Note that in the simplest case $\text{DOS}_{\mathbf{k}}(\omega) = 1$, and $D(\mathbf{k}) = f(-\infty) - f(\infty) = 1$, so we arrive back at the formula (5). In case of the pseudogap-modified spectra we, based on the experimental data, model DOS using the following function:

$$\text{DOS}_{\mathbf{k}}(\omega) = \begin{cases} 1, & |\omega| \geq 2\Delta; \\ |\omega|/2\Delta, & |\omega| < 2\Delta. \end{cases} \quad (8)$$

To obtain the experimental input on the electronic structure of $TaSe_2$ and $Cu_{0.2}NbS_2$ we have carried out a series of ARPES measurements for temperatures ranging from 300 down to 30 K, see Fig. 1 (a)–(b) and for further details Ref. 7. In the spectra of $TaSe_2$ the pseudogap is present already at the room temperature, and begins to increase sharply upon the transition into the incommensurate CDW state evolving to the band gap in the commensurate CDW state [7]. Magnitude of the pseudogap depends on the position in the Brillouin zone. In case of $TaSe_2$ the K-barrel is mostly affected by the pseudogap, so its contribution to the Hall coefficient has the strongest variation with temperature, which is the main reason for Hall coefficient to change sign. Fermi surface reconstruction also implies opening of the pseudogap on the parts of bone-shaped sheet around the M point and fading of the Γ-barrel ($\text{DOS}_{\mathbf{k}}(\omega) = \text{const} < 1$) near the point where it approaches the M-bone, see Fig. 1 (a)–(b) and ref. 7. In Fig.1 (e) contribution to the $\sigma_{xy}$ from different parts of the Fermi surface is shown for several temperatures. The above described procedure was also used in conjunction with $Cu_{0.2}NbS_2$ spectra [23]. As follows from the spectra, the electronic structure exhibits no considerable temperature dependence [Fig. 1 (c), (d), (f)], and is characterized by a uniform distribution of the spectral weigth. Calculated Hall coefficient of $NbS_2$ shows a weak temperature dependence. Comparing the result of the calculations with the experimental measurements we find a good agreement for the both studied compounds[Fig. 2], which implies correct implementation of the pseudogap effect into the semiclassical formulae.

In conclusion, we have shown that the suppression of the spectral weight at the Fermi level and its nonuniform distribution over the Fermi surface contours related to the pseudogap formation and consequent Fermi surface folding upon entering the CDW state are indispensable to attain quantitaive understanding of Hall coefficient temperature dependence in $TaSe_2$ and $Cu_{0.2}NbS_2$. Our findings hint that accounting for the pseudogap and Fermi surface reconstruction phenomenon may also be fruitful for understanding of other physical properties of CDW systems and

high-$T_c$ superconductors [3, 24].

The project is part of the FOR538 and was supported by the DFG under Grant No. KN393/4. We thank R. Hübel for technical support. ARPES experiments were performed using the "1$^3$ ARPES" end station at the UE112-lowE PGMa beamline of the Berliner Elektronenspeicherring-Gesellschaft für Synchrotron Strahlung m.b.H. (BESSY) and at the beamline SIS 9L, Swiss Light Source (SLS).

[1] H. N. S. Lee *et. al.*, J. Sol. State Chem. **1**, 190 (1970)
[2] M. Naito and S. Tanaka, J. Phys. Soc. Japan **51**, 219 (1982)
[3] D. LeBoeuf *et. al.*, Nature (London) **450**, 533 (2007)
[4] A. Kordyuk et. al., arXiv:0801.2546 (2008)
[5] T. Timisk *et. al.*, Rep. Prog. Phys. **62**, 61 (1999)
[6] H. Ding *et. al.*, Nature (London) **382**, 51 (1996)
[7] S. V. Borisenko *et. al.*, arXiv:0704.1544 (2007)
[8] Up to recent time band calculations failed to precisely reproduce the Fermi surface geometry of TaSe$_2$. Moreover, the evolution of electronic structure from the normal to CDW state still remains an open issue. These are actually the reasons why transport properties can not be calculated *a priori*.
[9] R. A. Craven, S. F. Meyer, Phys. Rev. B **16**, 4583 (1977)
[10] V. Vescoli, L. Degiorgi, H. Berger, L. Forro, Phys. Rev. Lett. **81**, 453 (1998)
[11] T. P. Beaulac, F. J. Pinski, P. B. Allen, Phys. Rev. B **23**, 3617 (1981)
[12] W. H. Butler, Phys. Rev. B **29**, 4224 (1984)
[13] N. P. Ong, Phys. Rev. B **43**, 193 (1991)
[14] P. T. Coleridge, Phys. Rev. B **44**, 3793 (1991)
[15] D. V. Evtushinsky *et. al.*, Phys. Rev. B **74**, 172509 (2006)
[16] Note that in the simplest case of electron-like round Fermi surface Eqs. (2) and (5) yield well known Drude formulae: $R_H = -1/ne$, $\sigma_{xx} = ne^2\tau/m$ where $n = (2\pi^2 L_c)^{-1} \int f(\epsilon(\mathbf{k})) \, d^2 \mathbf{k}$ and $m = \hbar k_F/v_F$.
[17] D. E. Moncton *et. al.*, Phys. Rev. B **16**, 801 (1977)
[18] A. Kordyuk *et. al.*, Phys. Rev. B **71**, 214513 (2005) V. B. Zabolotnyy *et. al.*, Phys. Rev. Lett. **96**, 037003 (2006)
[19] S. V. Borisenko *et. al.*, Phys. Rev. B **64**, 094513 (2001)
[20] In the case of spherically symmetric electronic structure, that is a perfect approximation for potassium, sodium and rubidium, one arrives at a simple formula for the Hall coefficient: $R_H = -3\pi^2 e^{-1} k_F^{-3}$, $k_F$ values for alkali metals can be found in literature [21]. Also in literature one can find calculations for metals with more complicated Fermi surfaces, e. g. copper and niobium [11]. Calculated Hall coefficients agrees with experimental within about 10%, as shown in the following table:

| $R_H$, $10^{-10}$ m$^3$/C | Na | K | Rb | Cu | Nb |
|---|---|---|---|---|---|
| Calculation | -2.38 | -4.49 | -5.4 | -0.530 | +0.752 |
| Experiment | -2.50 | -4.20 | -5.0 | -0.517 | +0.875 |

[21] A. vom Felde, J. Sprosser-Prou, J. Fink, Phys. Rev. B **40**, 10181 (1989) T. Schneider *et. al.*, Phys. Kondens. Materie **6**, 135 (1967)
[22] C. Kusko, R. S. Markiewicz, M. Lindroos, A. Bansil, Phys. Rev. B **66**, 140513(R) (2002)
[23] Intercalation of NbS$_2$ with 0.2 Cu affects only the size of Fermi surface barrels, while in general Fermi surface topology remains unchanged. Actually for pure NbS$_2$ we expect Hall coefficient to be roughly 20% less than for Cu$_{0.2}$NbS$_2$.
[24] S. Gnanarajan, R. F. Frindt, Phys. Rev. B **33**, 1443 (1986)